\documentclass[twocolumn,showpacs,preprintnumbers,amsmath,amssymb]{revtex4}

\usepackage{graphicx}
\usepackage{dcolumn}
\usepackage{bm}

\begin{document}

\title{Direct Observation of Long-Term Durability of Superconductivity in YBa$_2$Cu$_3$O$_7$-Ag$_2$O Composites}
\author{Juhn-Jong Lin,$^1${\footnote{E-mail: jjlin@mail.nctu.edu.tw}} Yong-Han Lin,$^1$ Shiu-Ming Huang,$^1$
Tsang-Chou Lee,$^1$ and Teng-Ming Chen$^2$}
\address{$^1$Institute of Physics, National Chiao Tung University, Hsinchu 300, Taiwan\\
$^2$Department of Applied Chemistry, National Chiao Tung University, Hsinchu 300, Taiwan}

\begin{abstract}

We report direct observation of long-term durability of superconductivity of several
YBa$_2$Cu$_3$O$_7$-Ag$_2$O composites that were first prepared and studied almost 14 years ago [J.
J.  Lin {\it et al}., Jpn. J. Appl. Phys. {\bf 29}, 497 (1990)]. Remeasurements performed recently
on both resistances and magnetizations indicate a sharp critical transition temperature at 91 K.
We also find that such long-term environmental stability of high-temperature superconductivity can
only be achieved in YBa$_2$Cu$_3$O$_7$ with Ag$_2$O addition, but not with pure Ag addition.

\end{abstract}
\maketitle

The science and technology of superconductivity has long been a fascinating research subject in
both academics and industry for almost a century now. In particular, the discoveries of
high-temperature superconductors in 1986 and in subsequent years, have inspired many scientists
and engineers to devote a tremendous amount of effort to the research and development of
large-scale as well as electronic applications of superconductivity [1-5]. However, it was soon
realized that advanced applications of high-temperature superconductors were hindered by the
absence of long-term durability of such ceramic materials [6-8]. In this work, we have
investigated the superconducting properties of several YBa$_2$Cu$_3$O$_7$-Ag$_2$O composites that
were first prepared and studied almost 14 years ago [9]. We demonstrate that robust, bulk
superconductivity with a sharp critical transition temperature, $T_c$, above 90 K was well
preserved in these 14-year-old composites. No special precautions have been taken in storing these
materials since they were made in 1989. Our results provide direct and strong evidence for
long-term environmental stability of superconductivity observed in the cuprate superconductors $-$
in particular in the most widely utilized YBa$_2$Cu$_3$O$_7$ compounds. This observation points to
an easy and reliable way for feasible fabrication of high-temperature superconductors for advanced
applications.

\begin{table}
\caption{\label{t1} Values for the relevant parameters of YBa$_2$Cu$_3$O$_7$-Ag$_2$O composites.
The values of $\rho$(300\,K) (in $\mu \Omega$ cm) and the Meissner fraction listed in front of the
parentheses were recently measured, while the values listed in the parentheses were those
originally reported in Ref. [9]. The Meissner fraction denotes the ratio of the field-cooled
magnetization to the zero-field-cooled magnetization at 5 K.}

\begin{ruledtabular}
\begin{tabular}{lccc}
Sample & weight ratio & $\rho$(300\,K) & Meissner fraction (\%) \\ \hline
\#9-1 & 10 & 3.4 (1.6) & 25 (31) \\
\#9-3 & 6 & 3.3 (1.7)  & 23 (22) \\
\#9-4 & 4 & 2.8 (2.5) & 21 (23)
\end{tabular}
\end{ruledtabular}
\end{table}

Our samples were fabricated in 1989 by the standard solid-state reaction method as previously
described in [9]. Appropriate amounts of stoichiometric YBa$_2$Cu$_3$O$_7$ and Ag$_2$O were mixed
and heated at 910$^\circ$C in air for 48 h. The amount of starting material was chosen so that,
after Ag$_2$O decomposed, the weight ratios of YBa$_2$Cu$_3$O$_7$ to Ag in the three samples were
equal to 10, 6, and 4. The fired mixture was then reground and pelletized. The pellets were then
fired at 935$^\circ$C for one day and subsequently were annealed at 470-480$^\circ$C in flowing
O$_2$ for 24 h and then slowly cooled to ambient in O$_2$. The resistances and magnetizations of
the samples were first reported in [9], where both measurements showed a sharp superconducting
transition with a $T_c \approx$ 91.5$\pm$0.5 K for all samples. After those measurements, the
samples were placed in an ordinary desiccator without vacuum for the last 14 years. The
desiccator, in turn, has been kept at room temperatures that vary approximately between
15$^\circ$C and 30$^\circ$C throughout the year. The relative humidity inside the desiccator
generally remains at around 35\% or slightly below 40\%. The values for the relevant parameters of
our samples are listed in Table I.

\begin{figure}
\includegraphics[angle=270,scale=0.3]{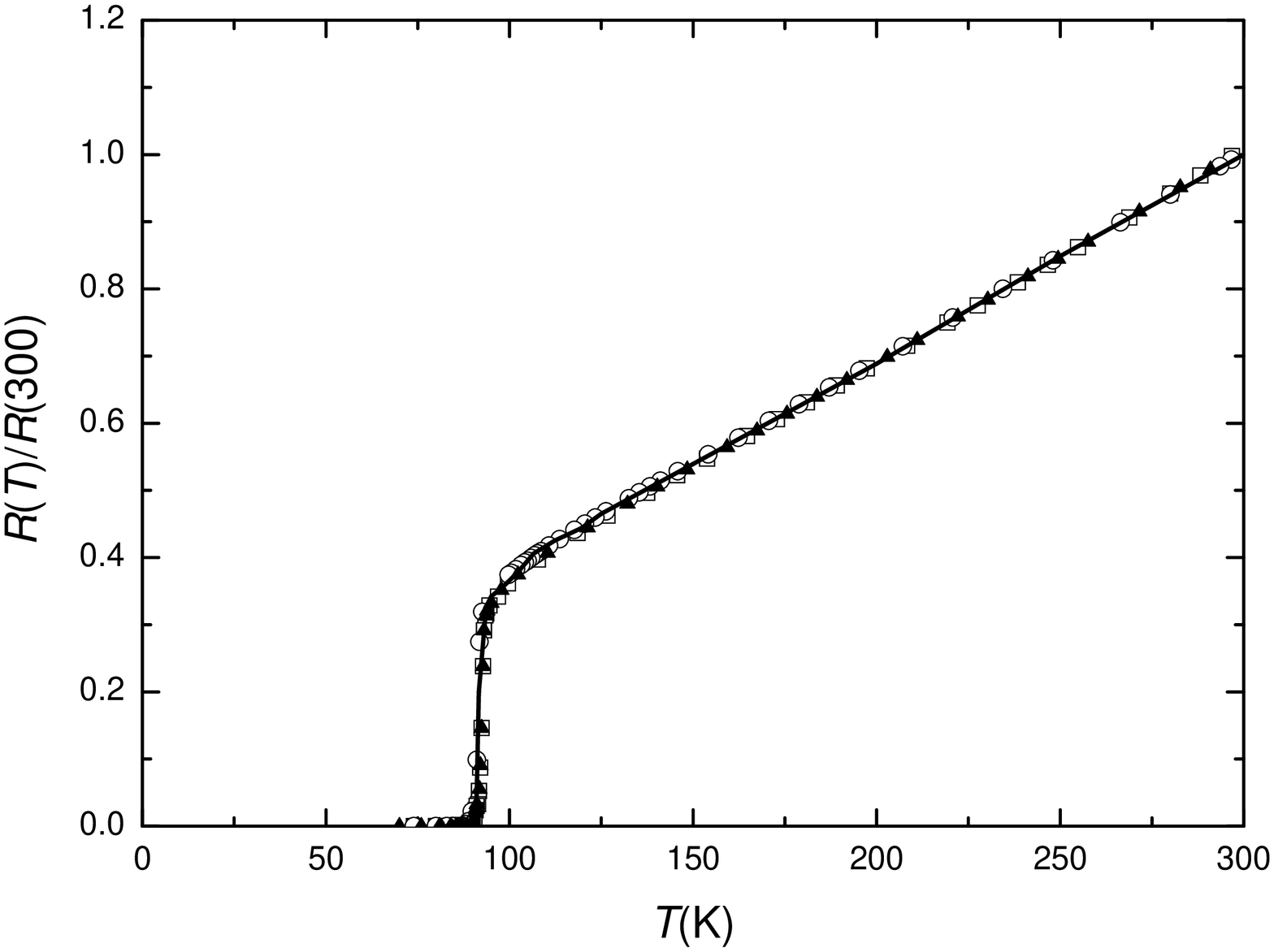}
\caption{\label{fig1} Normalized resistance $R(T)/R$(300\,K) as a function of temperature for
samples \#9-1 (squares), \#9-3 (circles), and \#9-4 (triangles) recently remeasured. The solid
line is taken from Ref. [9].}
\end{figure}

Figure 1 shows a plot of the normalized resistance, $R(T)/R$(300\,K), as a function of temperature
recently remeasured for these aged samples. The resistances (plotted as the solid curve) for these
samples first reported in [9] are also reproduced for comparison. It is clearly seen that the
normalized resistances for all samples are very similar. Remarkably, the new data and the old data
overlap and are practically indistinguishable. This result strongly demonstrates the absence of
any (noticeable) degradation in superconductivity in these materials for almost fourteen years.

Zero-field-cooled and field-cooled measurements (Fig. 2) performed recently on these aged samples
also confirm a very sharp superconducting transition with a $T_c$ similar to that determined from
resistance measurements. At 5 K, a large ratio ($>$ 20\%) of the field-cooled magnetization to the
zero-field-cooled magnetization is observed, clearly suggesting that the superconductivity in
these materials is maintained as a bulk property.

\begin{figure}
\includegraphics[angle=270,scale=0.3]{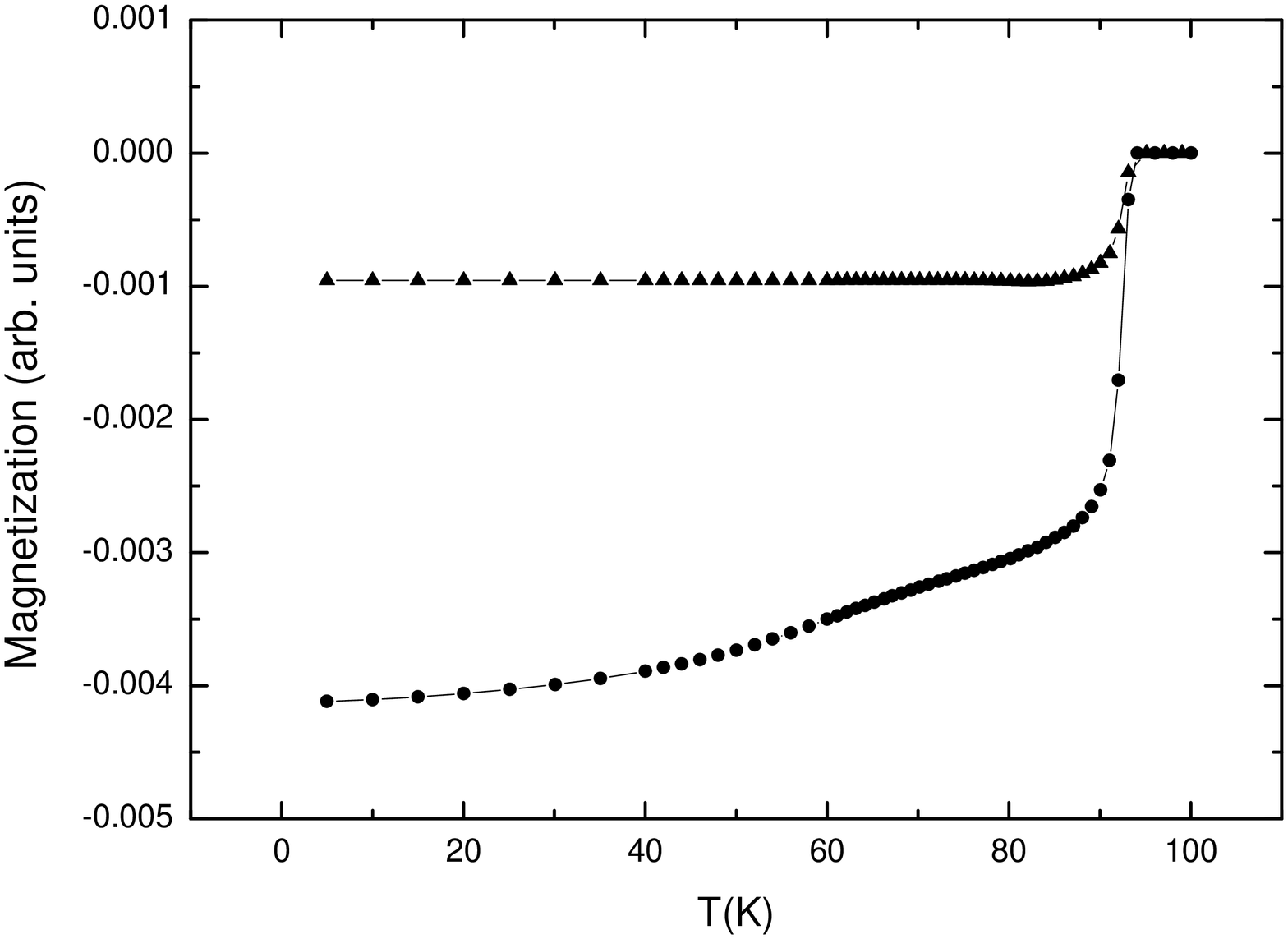}
\caption{\label{fig2} Zero-field-cooled (circles) and field-cooled (triangles) magnetization at 10
Oe for the aged sample \#9-3.}
\end{figure}

\begin{figure}
\includegraphics[angle=270,scale=0.3]{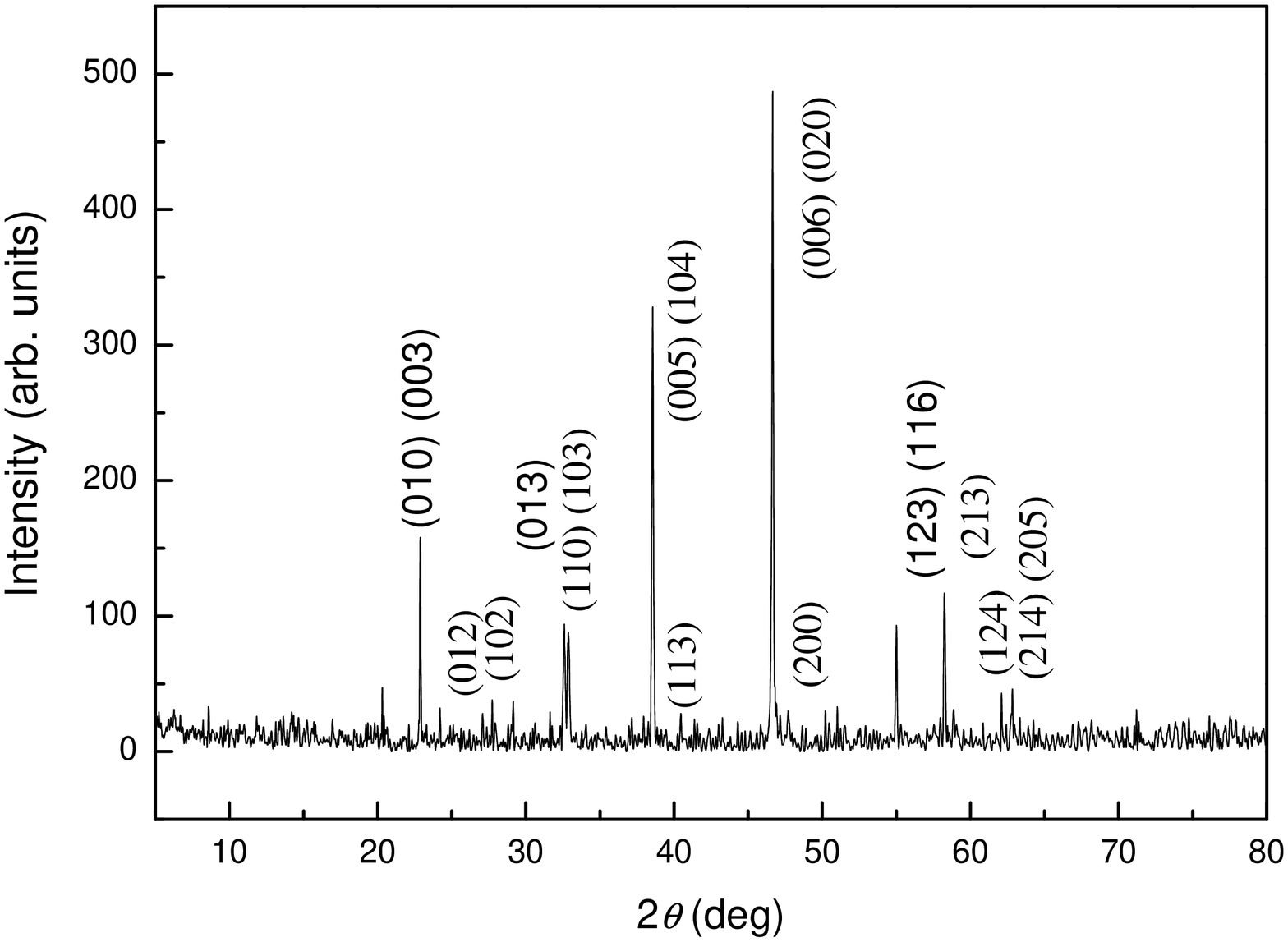}
\caption{\label{fig3} Powder x-ray diffraction pattern of the aged sample \#9-1.}
\end{figure}

X-ray diffraction data for these samples were recently remeasured and are shown in Fig. 3. The
diffraction pattern characteristic of YBa$_2$Cu$_3$O$_7$ is clearly revealed in these aged
samples. The diffraction peaks characteristic of Ag, which should be largest at 2$\theta$ =
38.1$^\circ$, 44.3$^\circ$, and 77.5$^\circ$, are barely seen in the figure.

Insofar as applications are concerned, our results indicate that only a small amount of Ag$_2$O
addition is sufficient to achieve long-term durability of superconductivity in the
YBa$_2$Cu$_3$O$_7$ compound. There are no noticeable differences in the electrical-transport,
magnetic, and structural properties in these composites with different weight ratios. This
observation points to an essential way of reducing the cost in the fabrication of this material
system for industrial applications.

To check the importance of Ag$_2$O addition in the fabrication of robust YBa$_2$Cu$_3$O$_7$ with
long-term environmental stability, we have also reexamined several aged YBa$_2$Cu$_3$O$_7$-Ag
composites (taken from [10]) that were made in a similar manner at about the same time as the
YBa$_2$Cu$_3$O$_7$-Ag$_2$O composites discussed above. While our YBa$_2$Cu$_3$O$_7$-Ag and
YBa$_2$Cu$_3$O$_7$-Ag$_2$O composites demonstrated essentially similar normal-state and
superconducting properties when they were freshly made [9,10], we now find absolutely no trace of
a resistive transition to a superconducting state down to 2 K in any of our aged
YBa$_2$Cu$_3$O$_7$-Ag composites. In addition, recent electrical-transport measurements revealed
much higher normal-state resistivities in YBa$_2$Cu$_3$O$_7$-Ag than in
YBa$_2$Cu$_3$O$_7$-Ag$_2$O, while magnetization measurements indicated a small Meissner fraction
of about 13\% in the aged YBa$_2$Cu$_3$O$_7$-Ag composites. These results suggest that Ag$_2$O,
but not pure Ag, should be added to achieve long-term durability of superconductivity in the
YBa$_2$Cu$_3$O$_7$ compound. Finally, we mention that most of our pure (i.e., Ag$_2$O- or Ag-free)
YBa$_2$Cu$_3$O$_7$ compounds prepared in the same time frame from 1989 to 1990 either are
decomposed now or show insulating behavior (i.e., an increasing resistance with decreasing
temperature) below room temperature.

The authors are grateful to D. C. Yan and C. C. Chi for magnetization measurements. This work was
supported by the Taiwan National Science Council through Grant No. NSC 91-2112-M-009-028.


\begin{flushleft}
{\bf References}
\end{flushleft}

\begin{enumerate}

\item S. Tanaka, Physica C {\bf 341}-{\bf 348}, 31 (2000).

\item D. Larbalestier, {\it et al.}, Nature {\bf 414}, 368 (2001).

\item D. Koelle, {\it et al.}, Rev. Mod. Phys. {\bf 71}, 631 (1999).

\item M. Nisenoff, and W. J. Meyers, IEEE T. Appl. Supercon. {\bf 11}, 799 (2001).

\item E. Polturak, {\it et al.} IEEE T. Microw. Theory \& Tech. 48, 1289, Part 2 (2000).

\item M. Regier, {\it et al.}, IEEE T. Appl. Supercon. {\bf 9}, 2375, Part 2 (1999).

\item J. P. Zhou, {\it et al.}, Physica C {\bf 273}, 223 (1997).

\item J. P. Zhou, {\it et al.}, J. Mater. Res. {\bf 12}, 2958 (1997).

\item J. J. Lin, {\it et al.}, Jpn. J. Appl. Phys. {\bf 29}, 497 (1990).

\item J. J. Lin and T. M. Chen, Z. Phys. B-Condens. Matter {\bf 81}, 13 (1990).

\end{enumerate}

\end{document}